\documentclass{article} 
\usepackage{amsmath,latexsym}
\usepackage{epstopdf,caption,subcaption,graphicx,xcolor,hyperref}
\usepackage{algpseudocode}
\usepackage{algorithm}
\usepackage{tikz}
\usetikzlibrary{shapes,arrows,fit,calc,positioning}
\usetikzlibrary{decorations.pathreplacing}
\normalsize
\newtheorem{theorem}{Theorem}
\newtheorem{lemma}{Lemma}
\newtheorem{operation}{Operation}

\newtheorem{definition}{Definition}
\newtheorem{example}{Example}

\newenvironment{proof}{{\sc Proof. }}{\hfill$\Box$\vspace{0.1in}}
\begin{document}

\title{Approximation algorithms for scheduling with rejection in green manufacturing}

\author{Mingyang Gong\footnote{Gianforte School of Computing, Montana State University, Bozeman, MT 59717, USA.
Email: {\tt mingyang.gong@montana.edu}}
\and
Brendan Mumey\footnote{Gianforte School of Computing, Montana State University, Bozeman, MT 59717, USA.
Email: {\tt brendan.mumey@montana.edu}}}

\date{}
\maketitle

\begin{abstract}
Motivated by green manufacturing, this paper investigates a scheduling with rejection problem subject to an energy consumption constraint.
Machines are associated with non-uniform energy consumption rates, defined as the energy consumed per unit time.
Each job is either rejected with a rejection penalty or accepted and scheduled on some machine for processing,
which incurs energy consumption.
The problem aims to minimize the makespan of the accepted jobs plus the total penalty of the rejected jobs while the total energy consumption is bounded by a given threshold.
In this paper,  when the number of machines is part of the input, we develop the first $(2+\epsilon)$-approximation algorithm for any fixed constant $\epsilon$ and a simple QPTAS
as well as a PTAS for uniform energy consumption rates.
Moreover, we present an FPTAS when the number of machines is a fixed constant.
\end{abstract}

\section{Introduction}\label{sec:intro}

With growing concerns about the scarcity of energy resources, energy consumption has become one of the key criteria in the modern production processes.
As a result, {\em green manufacturing}, an environmentally friendly approach, has been adopted by firms and governments, 
which takes both operational efficiency and energy consumption into consideration.
In other words, green manufacturing aims to optimize a certain objective under the constraint 
that the total energy consumption is no more than a predefined upper bound.
The energy consumption rate of a machine in manufacturing systems is defined as the amount of energy consumption per unit time on the machine.

{\em Scheduling with rejection}~\cite{BLM00} is another famous framework that captures many real-life scenarios.
In this framework, each job can be either accepted or rejected with a paid penalty for rejection.
Recently, Kong et al.~\cite{KSM25} studied a scheduling with rejection problem in green manufacturing assuming that the energy consumption rates are uniform.
The objective is to minimize the makespan of the accepted jobs plus the total rejection penalty of the rejected jobs,
with the total energy consumption limited by a given upper bound.
Our model extends the problem discussed in~\cite{KSM25} by allowing non-uniform energy consumption rates and
it is formally formulated in the next subsection.

\subsection{Model}

The problem studied in the paper can be described as follows.
An instance of our problem consists of a set of $m$ parallel machines $M = \{ M_1, \ldots, M_m \}$, a set of $n$ independent jobs $J = \{ J_1, \ldots, J_n \}$ and an upper bound $U > 0$.
Each machine $M_i$ has a non-uniform energy consumption rate $r_i$ and by reindexing the machines, we can assume $r_1 \le r_2 \le \ldots \le r_m$.
Each job $J_j$ is either rejected with a rejection penalty $e_j > 0$ or;
accepted and scheduled on one of the machines for processing with a processing time $p_j > 0$.
The energy consumed by $M_i$ is $r_i$ times the total processing time of the jobs scheduled on $M_i$.
The objective is to minimize the makespan of the accepted jobs plus the total rejection penalty of the rejected jobs
while ensuring that the total energy consumption is bounded by $U$.
In the three-field notation, we denote the problem by $P \mid rej, U, r_i  \mid C_{\max}(A) + \sum_{J_j \in R} e_j$,
where the number $m$ of machines is part of the input, ``$rej$'' indicates that jobs can be rejected;
$U$ is a predefined upper bound on the total energy consumption;
$r_i$ is the non-uniform energy consumption rate of $M_i$;
$A$ and $R$ are the sets of accepted and rejected jobs, respectively and
$C_{\max}(A)$ stands for the makespan of the schedule for $A$.
When $m$ is a fixed constant, we use ``$Pm$'' to replace ``$P$'' in the above notation.
We study the problem from the perspective of approximation algorithms.

Given a minimization problem, an {\em $\alpha$-approximation} algorithm is an algorithm
that produces a {\em feasible solution} for every instance $I$ in polynomial time in the input size $n$ of $I$
with an objective value of at most $\alpha$ times the optimal objective value.
A {\em quasi-polynomial time approximation scheme} or a {\em QPTAS} for short is a set of approximation algorithms
such that given a fixed $0 < \epsilon \le 1$, it contains a $(1+\epsilon)$-approximation algorithm that runs in $n^{(\log n)^{O(1)}}$ time.
When the running time is limited to a polynomial in $n$ for a fixed $\epsilon$, then a QPTAS becomes a {\em polynomial-time approximation scheme} or a {\em PTAS} for short.
Similarly, a PTAS becomes an {\em efficient} PTAS (EPTAS for short) if the algorithm runs in $f(\frac 1\epsilon) n^c$ where $c$ is a constant and $f$ is a polynomial-time computable function.
An EPTAS becomes a {\em fully} PTAS (FPTAS) if $f$ is a polynomial both in $n$ and $\frac 1\epsilon$.

\subsection{Literature review for related problems}

When the energy consumption rates are uniform, the energy consumption constraint is equivalent to saying that the total processing time of the accepted jobs does not exceed a given threshold.
Kong et al.~\cite{KSM25} proposed the first $2$-approximation algorithm for $P \mid rej, U, r_i = r \mid C_{\max}(A)+\sum_{J_j \in R} e_j$ where ``$r_j = r$'' represents that the energy consumption rates are uniform.

Recall that $r_1 \le r_2 \le \ldots \le r_m$.
When $r_1 \sum_j p_j \le U$, we allow to accept all jobs because they can be scheduled on $M_1$ to satisfy the energy consumption constraint.
Therefore, if $e_j = \infty$ for each $J_j$ and $r_1 \sum_j p_j \le U$, then the scheduler is forced to accept all jobs and the problem is reduced to $P \mid U, r_i \mid C_{\max}$.
Li et al.~\cite{LZL16} initialized the study of $P \mid U, r_i \mid C_{\max}$ and presented a $2$-approximation algorithm.
Subsequently, Jiang et al.~\cite{JTL23} developed an improved algorithm with an approximation ratio of $\frac {\sqrt{33}+1}4 \approx 1.686$.
Finally, Li and Ou~\cite{LO24} designed an improved $(\frac 43 + \epsilon)$-approximation algorithm for any fixed $\epsilon > 0$ 
and finally solved this problem by achieving a novel PTAS.
Moreover, they also developed an FPTAS for $Pm \mid U, r_i \mid C_{\max}$ where the number $m$ of machines is a constant.
One clearly sees that once the set $A$ of accepted jobs is determined, given $\epsilon>0$, we can apply the PTAS in~\cite{LO24} for $P \mid U, r_i \mid C_{\max}$ to obtain a schedule for $A$
such that the energy consumption constraint is satisfied and the makespan is at most $1+\epsilon$ times the minimum makespan of $A$.
Therefore, we use this PTAS as a subroutine of our algorithms as defined in Definition~\ref{def01}.

When $U = \infty$, we can ignore the energy consumption constraint since it is always satisfied.
Therefore, the problem is reduced to the classic scheduling with rejection problem $P \mid rej \mid C_{\max}(A)+\sum_{J_j \in R} e_j$.
Bartal et al. \cite{BLM00} pioneered the research for the scheduling with rejection problem and 
designed a PTAS for a general $m$ and an FPTAS for a fixed $m$, respectively.
Kones et al. \cite{KL19} developed an EPTAS for scheduling with rejection and several related problems using a general framework.

If $U = \infty$ and $e_j = \infty$ for each job $J_j$, then the problem becomes the well-known multiprocessor scheduling problem $P \mid \mid C_{\max}$~\cite{GJ79}.
Hochbaum and Shmoys~\cite{HS87} showed that $P \mid \mid C_{\max}$ is strongly NP-hard and achieved a PTAS.
Before the PTAS, Graham~\cite{Gra66} presented the famous {\em list scheduling}~\cite{Gra66}
and {\em longest processing time first} algorithm~\cite{Gra69},
which achieves a tight approximation ratio of $2-\frac 1m$ and $\frac 43-\frac 1{3m}$, respectively,
where $m$ is the number of machines.
When $m \ge 2$ is a fixed constant, $Pm \mid \mid C_{\max}$ is weakly NP-hard~\cite{GJ79}, for which Sahni presented an FPTAS~\cite{Sah76}.

\subsection{Our results}

We develop the first $(2+\epsilon)$-approximation algorithm for the problem $P \mid rej, U, r_i=r  \mid C_{\max}(A) + \sum_{J_j \in R} e_j$ for any fixed $\epsilon > 0$
and then, present a QPTAS where the number $m$ of machines is part of the input.
In the $(2+\epsilon)$-approximation algorithm, we first guess the value $Q^*$ in Eq.(\ref{eq01}), 
which is the maximum of the longest processing time among the accepted jobs and
the largest rejection penalty among the rejected jobs in an optimal solution.
We then partition the jobs into {\em $\gamma$-critical} and {\em $\gamma$-tiny} as in Definition~\ref{def02} 
where a job $J_j$ is $\gamma$-critical if $e_j > \gamma Q^*$ with $\gamma = \frac \epsilon6$.
The rejection decisions for the $\gamma$-critical jobs can always be made easily, while
a linear program is formulated to make the rejection decisions for the $\gamma$-tiny jobs. 
Finally, we apply the PTAS for $P \mid U, r_i  \mid C_{\max}$~\cite{LO24} to schedule the accepted jobs.
Different from the $(2+\epsilon)$-approximation, we set $\gamma = \frac {3\epsilon}{4n}$ and all $\gamma$-tiny jobs are simply rejected in our QPTAS.
When the energy consumption rates are uniform, we achieve a PTAS in which the $\gamma$-tiny jobs are further partitioned into {\em long} and {\em short} and 
the rejection decisions for long jobs are carefully determined by the linear program.
Lastly, we present an FPTAS using the so-called ``vector trimming'' technique in~\cite{W00} when the number $m$ of machines is a fixed constant.

The remainder of this paper is organized as follows.
In Section~2, we present some definitions and notations related to the problem.
Section~3 shows how to handle the $\gamma$-critical jobs.
In Sections~4 and~5, we describe our $(2+\epsilon)$-approximation algorithm and QPTAS for $P \mid rej, \hat{U}, r_i \mid C_{\max}(A)+\sum_{J_j \in R} e_j$, respectively.
Section~6 provides a PTAS for uniform energy consumption rates while 
Section~7 presents an FPTAS when $m$ is a fixed constant.
Finally, Section~8 summarizes the paper and provides some possible future research problems.

\section{Preliminaries}
\label{sec2}

We introduce the notations used in this paper as follows.
Suppose that we are given a set of $m$ parallel machines $M = \{ M_1, \ldots, M_m \}$, a set of $n$ independent jobs $J = \{ J_1, \ldots, J_n \}$ and an upper bound $U > 0$ on the total enery consumption.
Each machine $M_i$ is associated with an energy consumption rate $r_i$ and 
without loss of generality, we assume $0 \le r_1 \le r_2 \le \ldots \le r_m$.
Each job $J_j$ has a processing time $p_j > 0$ and a rejection penality $e_j > 0$.
The scheduler should decide to reject $J_j$ or not and if $J_j$ is accepted, then $J_j$ will be scheduled on some machine for processing.
For every $i = 1, \ldots, m$, the {\em load} $t_i$ of $M_i$ is the total processing time of the jobs scheduled on $M_i$.
Given a schedule, the total enery consumption is $\sum_{i=1}^m r_i t_i$ and the makespan is the maximum load of machines.
The problem aims to minimize the makespan of the accepted jobs plus the total penalty of the rejected jobs while the total energy consumption does not exceed $U$, which is referred to as the ``energy consumption constraint'' in the later context.
In the three-field notation, the problem is denoted by $P \mid rej, U, r_i  \mid C_{\max}(A) + \sum_{J_j \in R} e_j$.

Given an instance $I$ of $P \mid rej, U, r_i  \mid C_{\max}(A) + \sum_{J_j \in R} e_j$, a {\em feasible solution} for $I$ consists of two parts:
\begin{itemize}
\item A set $R \subseteq J$ of rejected jobs;

\item A schedule $\pi$ for the set $A = J \setminus R$ of accepted jobs such that the total energy consumption is at most $U$.
\end{itemize}
The {\em cost} of a feasible solution is the makespan of $\pi$ plus the total rejection penalty of the rejected jobs.
Note that a feasible solution must exist since all jobs can be rejected.
Denote by $Z^*$ the minimum cost among all feasible solutions for $I$ and 
an {\em optimal solution} is a feasible solution with cost $Z^*$.
We fix an optimal solution for $I$ for discussion.
Let $A^*$ and $R^*$ be the sets of accepted and rejected jobs in the optimal solution, respectively
and let $\pi^*$ be the schedule for $A^*$ in the optimal solution.
For ease of presentation, given a subset $J' \subseteq J$ of jobs, we let $P(J')$ and $E(J')$ 
be the total processing time and total rejection penalty of the jobs in $J'$, respectively.
Similarly, denote by $p_{\max}(J')$ and $e_{\max}(J')$ the longest processing time and largest rejection penalty of the jobs in $J'$, respectively.
When $J' = \emptyset$, we set all $P(J')$, $E(J')$, $p_{\max}(J')$ and $e_{\max}(J')$ to be $0$. 
Moreover, given a schedule $\pi$ of $J'$, we let $C_{\max}(\pi)$ denote the makespan of $\pi$.
We remind the readers that in the three-field notation, although we use $C_{\max}(A)$ for consistency with the reference~\cite{KSM25}, it indeed should be $C_{\max}(\pi)$ where $\pi$ is a schedule for $A$.

When the set of accepted jobs is fixed, the problem is reduced to $P \mid U, r_i \mid C_{\max}$
which aims to find a schedule such that the makespan is minimized and the total energy consumption is at most $U$. 
The authors in~\cite{LO24} presented a PTAS for $P \mid U, r_i \mid C_{\max}$
and we use this PTAS as a subroutine in our algorithms.

\begin{definition}
\label{def01}
{\em (A PTAS $\Phi$ for $P \mid U, r_i \mid C_{\max}$~\cite{LO24})}
The authors in~\cite{LO24} presented a PTAS $\Phi$ for $P \mid U, r_i \mid C_{\max}$.
For ease of presentation, given an instance $I$ of $P \mid rej, U, r_i \mid C_{\max}(A)+\sum_{J_j \in R} e_j$ and a fixed constant $0 < \epsilon \le 1$, 
if $A$ is the set of accepted jobs, then we let $\Phi(A, \epsilon)$ denote the schedule produced by the PTAS $\Phi$ on $A$ by using $\epsilon$ as input.
\end{definition}

Next, we present a technique lemma which plays an important role in the later proofs.
Recall that the number of machines is $m$ and $r_i$ is the energy consumption rate of the machine $M_i$.

\begin{lemma}
\label{lemma01}
Given $x_1, x_2, \ldots, x_m \ge 0$ and $y_1, y_2, \ldots, y_m \ge 0$ with $\sum_{i=1}^m x_i \ge \sum_{i=1}^m y_i$,
if there exists a non-negative integer $\ell$ such that $y_i > x_i$ if and only if $i \le \ell$, then we have
\[
\sum_{i=1}^m r_i x_i \ge \sum_{i=1}^m r_i y_i.
\]
\end{lemma}
\begin{proof}
If $\ell = 0$, then $y_i \le x_i$ for every $i=1, \ldots, m$.
The lemma clearly holds.

Next, we assume $\ell \ge 1$.
If $\ell \ge m$, then $y_i > x_i$ for every $i=1, \ldots, m$ and thus, $\sum_{i=1}^m y_i > \sum_{i=1}^m x_i$, a contradiction.
So, $1 \le \ell \le m-1$.
Note that $y_i > x_i$ for every $i \le \ell$ and $y_i \le x_i$ for every $i \ge \ell+1$.
Therefore, by $0 \le r_1 \le \ldots \le r_m$ and $\sum_{i=1}^m x_i \ge \sum_{i=1}^m y_i$, we have
\begin{eqnarray*}
\sum_{i=1}^m r_i x_i - \sum_{i=1}^m r_i y_i  & = & \sum_{i=1}^\ell r_i (x_i-y_i) + \sum_{i=\ell+1}^m r_i (x_i-y_i) \\
& \ge & r_\ell \sum_{i=1}^\ell (x_i-y_i) + r_\ell \sum_{i=\ell+1}^m (x_i-y_i) \\
& = & r_\ell \sum_{i=1}^m (x_i-y_i) \ge 0.
\end{eqnarray*}
The lemma is proved.
\end{proof}

Consider a set $A$ of accepted jobs, for which there exists a schedule $\pi$ with makespan $C_{\max}(\pi)$
that satisfies the energy consumption constraint.

\begin{lemma}
\label{lemma02}
Let $A'$ be another set of accepted jobs such that $p_{\max}(A) \ge p_{\max}(A')$ and $P(A) \ge P(A')$.
Then, we can construct a schedule for $A'$ such that the energy consumption constraint is satisfied and the makespan is at most $2C_{\max}(\pi)$.
\end{lemma}
\begin{proof}
Consider the schedule $\pi$ for $A$ and we let $x_i$ be the load of $M_i$ in $\pi$.
Since $\pi$ satisfies the energy consumption constraint and has makespan $C_{\max}(\pi)$, we have $\sum_{i=1}^m r_i x_i \le U$ and $C_{\max}(\pi)=\max_{i=1}^m x_i \ge p_{\max}(A)$.

We construct a schedule $\pi'$ for $A'$ as follows: 
Consider the machines $M_1, \ldots, M_m$ in order.
For each machine $M_i$, we greedily assign the jobs in $A'$ to $M_i$ until the load first exceeds $x_i$ or we run out of jobs.
When some jobs remain unassigned, we proceed to the next machine $M_{i+1}$.
If the last machine $M_m$ is reached and some jobs are still unassigned, then we schedule all of them on $M_m$.

Next, we prove that $\pi'$ is the desired schedule.
Let $y_i$ be the load of $M_i$ in $\pi'$.
Therefore, the total energy consumption is $\sum_{i=1}^m r_iy_i$ and the makespan of $\pi'$ is $\max_{i=1}^m y_i$.
To finish the proof, it suffices to show $\sum_{i=1}^m r_i y_i \le U$ and $\max_{i=1}^m y_i \le 2C_{\max}(\pi)$.

We first show $\sum_{i=1}^m r_i y_i \le U$.
Note that $P(A) \ge P(A')$ is equivalent to $\sum_{i=1}^m x_i \ge \sum_{i=1}^m y_i$.
Let $\ell$ be the largest index such that $y_i > x_i$ for every $i \le \ell$.
If $y_i \le x_i$ for every $i = 1, \ldots, m$, then we set $\ell = 0$.
By $\sum_{i=1}^m x_i \ge \sum_{i=1}^m y_i$, one clearly sees that $\ell \le m-1$.
It follows from the definition of $\ell$ and $\ell \le m-1$ that $y_{\ell+1} \le x_{\ell+1}$.
Therefore, by the construction of $\pi'$, all jobs have been assigned after scheduling the jobs on $M_{\ell+1}$.
In other words, $y_i = 0$ for every $i = \ell+2, \ldots, m$.
So, the integer $\ell$ satisfies the property stated in Lemma~\ref{lemma01}.
By Lemma~\ref{lemma01} and $\sum_{i=1}^m r_i x_i \le U$, we have shown that $\sum_{i=1}^m r_iy_i \le \sum_{i=1}^m r_i x_i \le U$. 

Then, we show $\max_{i=1}^m y_i \le 2C_{\max}(\pi)$. 
We continue to use the integer $\ell$ as defined in the last paragraph.
Recall that $\ell \le m-1$ and so, $y_m \le x_m$.
By the construction of $\pi'$, one sees that $y_i \le x_i+p_{\max}(A')$ for every $i \le m-1$.
Using $C_{\max}(\pi)=\max_{i=1}^m x_i \ge p_{\max}(A)$ and $p_{\max}(A) \le p_{\max}(A')$, we have 
\[
C_{\max}(\pi') = \max_{i=1}^m y_i \le \max_{i=1}^m x_i + p_{\max}(A') 
\le C_{\max}(\pi) + p_{\max}(A) \le 2C_{\max}(\pi).
\]
The lemma is proved.
\end{proof}

Recall that $e_{\max}(R^*)$ is the largest penalty of the jobs in $R^*$ and $p_{\max}(A^*)$ is the longest processing time of the jobs in $A^*$, 
where $A^*$ and $R^*$ is the set of accepted and rejected jobs in the optimal schedule, respectively.
Since $e_{\max}(R^*) \in \{ 0, e_1, \ldots, e_n \}$ and $p_{\max}(A^*) \in \{ 0, p_1, \ldots, p_n \}$, the values of $e_{\max}(R^*)$ and $p_{\max}(A^*)$ can be guessed in $O(n^2)$ time.
In the sequel, we assume $e_{\max}(R^*)$ and $p_{\max}(A^*)$ are known and for ease of presentation, we let 
\begin{equation}
\label{eq01}
Q^* = \max \left \{ e_{\max}(R^*), p_{\max}(A^*) \right \} \le Z^*.
\end{equation}
Given a value $\gamma$, we distinguish the {\em $\gamma$-critical} and {\em $\gamma$-tiny} jobs as in Definition~\ref{def02}
and in the next section, we will show that the rejection decisions for the $\gamma$-critical jobs can be made easily.

\begin{definition}
\label{def02}
Given a value $\gamma > 0$, a job $J_j$ is {\em $\gamma$-critical} if $e_j > \gamma Q^*$; otherwise, $J_j$ is {\em $\gamma$-tiny}.
We let $C$ and $T$ denote the set of $\gamma$-critical and $\gamma$-tiny jobs, respectively.
\end{definition}

\section{Determining rejection decisions for $\gamma$-critical jobs}

In this section, we assume the value $\gamma$ is given and 
we will show how to determine the rejection decisions for the $\gamma$-critical jobs.
Let $C_0 = \{ J_j \in C: e_j > Q^* \}$.
For a $\gamma$-critical job $J_j \in C \setminus C_0$, by Definition~\ref{def02}, we have $\gamma Q^* < e_j \le Q^*$.
In other words, there exists an integer $1 \le i \le \log_{1+\epsilon/4} \frac 1\gamma$ such that 
$(1+\epsilon/4)^{-i} Q^* < e_j \le (1+\epsilon/4)^{1-i}  Q^*$.
Therefore, the $\gamma$-critical jobs in $C \setminus C_0$ can be partitioned into the following subgroups.
\begin{equation}
\label{eq02}
C_i = \left \{ J_j \in C \setminus C_0: (1 + \frac \epsilon4)^{-i} Q^* < e_j \le (1+\frac \epsilon4)^{1-i}  Q^* \right \},
\mbox{ for each integer $1 \le i \le \log_{1+\epsilon/4} \frac 1\gamma$.}
\end{equation}
The following lemma estimates the number of subgroups $C_i$.

\begin{lemma}
\label{lemma03}
The total number of subgroups $C_i$ is bounded by $O(\frac {\log 1/\gamma}\epsilon)$.
Moreover, if $J_j, J_k \in C_i$ for some $1 \le i \le \log_{1+\epsilon/4} \frac 1\gamma$, then $e_j \le (1+\frac \epsilon4)e_k$.
\end{lemma}
\begin{proof}
The total number of subgroups $C_i$ is at most $O(\log_{1+\epsilon/4} \frac 1\gamma) = O(\frac {\log 1/\gamma}\epsilon)$.
If $J_j, J_k \in C_i$, then by Eq.(\ref{eq02}), 
we have $e_j \le (1+\frac \epsilon4)^{1-i}  Q^*$ and $e_k \ge (1+\frac \epsilon4)^{-i}  Q^*$.
The proof is straightforward.
\end{proof}


The next definition describes the number $c^*_i$ of rejected jobs in each subgroup $C_i$ in the optimal solution
and clearly, $c^*_i \in \{ 0, \ldots, n \}$ for every $i$.

\begin{definition}
\label{def03}
For each non-negative integer $i$, let $c^*_i$ be the number of jobs in $C_i$ that are rejected in the optimal solution.
Note that $c^*_i \in \{ 0, \ldots, n \}$ has at most $n+1$ possible values.
By Lemma~\ref{lemma03}, all $c^*_i$ can be guessed in $(n+1)^{O(\frac {\log 1/\gamma}\epsilon)}$ time.

Recall that each job $J_j \in C_0$ has $e_j > Q^* \ge e_{\max}(R^*)$.
Therefore, no jobs in $C_0$ can be rejected in the optimal solution, that is, $c^*_0 = 0$.
\end{definition}
Once $c^*_i$ is known, we sort the jobs in $C_i$ in the non-increasing order of their processing times
and reject the first $c^*_i$ jobs in $C_i$.
We conclude this operation as follows, which determines the rejection decisions for all $\gamma$-critical jobs.

\begin{operation}
\label{op01}
Given a $\gamma$, we first partition the $\gamma$-critical jobs into $C_0$ and the subgroups $C_i$ as in Eq.(\ref{eq02}).
For each $C_i$, we guess the number $c^*_i$ as defined in Definition~\ref{def03}, which can be done in $(n+1)^{O(\frac {\log 1/\gamma}\epsilon)}$ time.
Subsequently, we sort the jobs in $C_i$ in the non-increasing order of their processing times 
and reject the first $c^*_i$ jobs in $C_i$.
\end{operation}

Let $A_c$ and $R_c$ be the set of critical jobs that are accepted and rejected by Operation~\ref{op01}, respectively.
So, $C = A_c \cup R_c$.
To describe the relationship between the job sets $A_c$ and $A^* \cap C$, we present the following definition
and we will prove that $A^* \cap C$ {\em covers} $A_c$ in the proof of Lemma~\ref{lemma04}.

\begin{definition}
\label{def04}
Given two sets of jobs $J', J'' \subseteq J$, we say $J''$ {\em covers} $J'$
if there exists an injective function $f: J' \to J''$ mapping every job $J_j \in J'$ to a job $J_{f(j)} \in J''$ such that $p_j \le p_{f(j)}$
where $p_j$ and $p_{f(j)}$ are the processing times of $J_j$ and $J_{f(j)}$, respectively.
Specifically, every job set covers an empty set.
\end{definition}
We give a simple example to explain Definition~\ref{def04}.
Note that $J', J''$ may not be disjoint when $J''$ covers $J'$ by Example~\ref{ex01}.

\begin{example}
\label{ex01}
Suppose that the processing times of the jobs in $J$ can be listed as $\langle 5, 4, 4, 3, 2, 1 \rangle$.
The set $J' \subseteq J$ consists of the last three jobs in $J$.
In other words, $J' = \{ J_4, J_5, J_6 \}$.
Clearly, $J'' = \{ J_1, J_2, J_3, J_4 \}$ covers $J'$ by mapping each job $J_j \in J'$ to $J_{j-3} \in J''$.
\end{example}

\begin{lemma}
\label{lemma04}
$A^* \cap C$ covers $A_c$.
\end{lemma}
\begin{proof}
It suffices to show that for each $0 \le i \le \log_{1+\epsilon/4} \frac 1\gamma$, $A^* \cap C_i$ covers $A_c \cap C_i$.
Note that the jobs in $C_i$ are sorted in the non-increasing order of their processing times
and by Operation~\ref{op01} and Definition~\ref{def03}, $|A^* \cap C_i| = |A_c \cap C_i| = |C_i| - c^*_i$.

We design a function $f$ from $A_c \cap C_i$ to $A^* \cap C_i$ as follows:
We pick the job in $A_c \cap C_i$ with the smallest job index and map it to the job in $A^* \cap C_i$ with the smallest job index.
Afterwards, we remove these two jobs from $A_c \cap C_i$ and $A^* \cap C_i$ and $|A_c \cap C_i| = |A^* \cap C_i|$ still holds.
We repeatedly do the above step until every job in $A_c \cap C_i$ has been mapped to some job in $A^* \cap C_i$.
By Operation~\ref{op01}, $A_c \cap C_i$ consists of $|C_i| - c^*_i$ jobs in $C_i$ with the largest job indices.
So, for a job $J_j \in A_c \cap C_i$, $f(j) \le j$ and by the sorting rule of $C_i$, we have $p_{f(j)} \ge p_j$.
The proof is done.
\end{proof}

Next, we estimate the total rejection penalty of the jobs in $R_c$.


\begin{lemma}
\label{lemma05}
$P(A_c) \le P(A^* \cap C)$ and $p_{\max}(A_c) \le p_{\max}(A^* \cap C)$.
Moreover, $E(R_c) \le (1+\frac \epsilon4) E(R^* \cap C)$.
\end{lemma}
\begin{proof}
By Lemma~\ref{lemma04}, $A^* \cap C$ covers $A_c$.
The first two inequalities are straightforward from Definition~\ref{def04}.

By Definition~\ref{def03}, $c^*_0 = 0$ and thus, $R_c \cap C_0 = \emptyset$.
Therefore, it suffices to show $E(R_c \cap C_i) \le (1+\frac \epsilon4) E(R^* \cap C_i)$ for every $1 \le i \le \log_{1+\epsilon/4} \frac 1\gamma$.
By Definition~\ref{def03} and Operation~\ref{op01}, we know $|R^* \cap C_i| = |R_c \cap C_i| = c^*_i$.
The lemma is proved by Lemma~\ref{lemma03}.
\end{proof}

\section{A $(2+\epsilon)$-approximation for $P \mid rej, U, r_i  \mid C_{\max}(A) + \sum_{J_j \in R} e_j$}

We assume that $0<\epsilon \le 1$ is a fixed constant and for simplicity, we further assume $\frac 1\epsilon$ is an integer.
In this section, given $\epsilon$, we present a $(2+\epsilon)$-approximation algorithm for $P \mid rej, U, r_i  \mid C_{\max}(A) + \sum_{J_j \in R} e_j$ where the number $m$ of machines is part of the input.
We set $\gamma = \frac \epsilon 6$ and we apply Operation~\ref{op01} to determine all $\gamma$-critical jobs.
Note that the values $c^*_i$ in Definition~\ref{def03} can be guessed in $(n+1)^{O(1/\epsilon^2)}$, which is polynomial in $n$ for any fixed $\epsilon$.
So, Operation~\ref{op01} can terminate in polynomial time.

\subsection{Determining rejection decisions for $\gamma$-tiny jobs}

In this subsection, we present the operation that makes the rejection decision for each $\gamma$-tiny job in $T$.
By Eq.(\ref{eq01}), we have $E(R^* \cap T) \le n \cdot e_{\max}(R^*) \le n \cdot Q^*$.
Recall that $\frac 1{2\gamma} = \frac 3\epsilon$ is an integer and thus, $\lfloor \frac {E(R^* \cap T)}{2\gamma Q^*} \rfloor \le \frac n {2\gamma}$.

\begin{definition}
\label{def05}
We let $t^* = \lfloor \frac {E(R^* \cap T)}{2\gamma Q^*} \rfloor$ and we know $t^* \in \{ 0, \ldots, \frac n {2\gamma} \}$.
Clearly, $t^* \cdot 2\gamma Q^* \le E(R^* \cap T) \le (t^*+1) \cdot 2\gamma Q^*$.
\end{definition}
Since $t^*$ has at most $1+\frac n{2\gamma}$ possible values, we can guess the value of $t^*$ in $O(\frac n\epsilon)$ time.
Subsequently, we establish a linear program LP(\ref{eq03}) to make rejection decisions for all $\gamma$-tiny jobs.
For each $\gamma$-tiny job $J_j \in T$, we let $x_j, y_j \in \{ 0, 1 \}$ be two binary decision variables such that $x_j + y_j = 1$
and $x_j = 1$ if and only if $J_j$ is rejected in the optimal solution.
If $p_j > p_{\max}(A^*)$, then obviously, $J_j$ is rejected in the optimal solution and thus, we let $x_j = 1$ immediately.
We relax the variables $x_j, y_j$ to be continuous in LP(\ref{eq03}).
One sees that LP(\ref{eq03}) minimizes the total processing time of the accepted jobs in $T$
while the total penalty of the rejected jobs in $T$ is bounded by  $(t^*+1) \cdot 2\gamma Q^*$.

\begin{equation}
\label{eq03}
\begin{aligned}
 \min \quad & \sum_{J_j \in T} y_j p_j   \\
& x_j = 1 \mbox{ and } y_j = 0, \quad  \forall J_j  \in T \mbox{ and } p_j > p_{\max}(A^*) \\
& x_j + y_j = 1, \quad  \forall J_j  \in T \mbox{ and } p_j \le p_{\max}(A^*) \\
& \sum_{J_j \in T} x_j e_j  \le (t^*+1)\cdot 2\gamma Q^*.
\end{aligned}
\end{equation}
Note that LP(\ref{eq03}) admits a feasible solution by setting $x_j = 1$ if and only if $J_j \in R^* \cap T$.
We compute a basic solution for LP(\ref{eq03}) and partition the $\gamma$-tiny jobs into three sets.
\begin{equation}
\label{eq04}
A_t = \{ J_j \in T: y_j = 1 \},
T_1 = \{ J_j \in T: x_j = 1 \} \mbox{ and }
T_2 = T \setminus \{ A_t \cup T_1 \}.
\end{equation}
The jobs in $A_t$ are accepted while the other $\gamma$-tiny jobs are rejected.
For simplicity, we let $R_t = T_1 \cup T_2$.
The operation is summarized as follows.

\begin{operation}
\label{op02}
We first guess the value $t^*$ as defined in Definition~\ref{def05} and establish the linear program LP(\ref{eq03}).
Then, we compute a basic solution and partition the $\gamma$-tiny jobs into $A_t$ and $R_t$ as defined in Eq.(\ref{eq04}).
The jobs in $A_t$ are accepted while the jobs in $R_t = T_1 \cup T_2$ are rejected.
\end{operation}

The next lemma shows that $T_2$ consists of at most one job.

\begin{lemma}
\label{lemma06}
$|T_2| \le 1$.
\end{lemma}
\begin{proof}
Note that LP(\ref{eq03}) has $|T|+1$ non-trivial constraints and thus, in a basic solution, there exist at most $|T|+1$ positive values.
Clearly, by Eq.(\ref{eq04}), the number of positive values is exactly $|A_t|+|T_1|+2|T_2|=|T|+|T_2|$.
Therefore, $|T|+|T_2| \le |T|+1$ and we have $|T_2| \le 1$.
\end{proof}

Now, we are ready to estimate the values $P(A_t)$ and  $E(R_t)$.

\begin{lemma}
\label{lemma07}
$P(A_t) \le P(A^* \cap T)$ and $p_{\max}(A_t) \le p_{\max}(A^*)$.
Moreover, $E(R_t) \le E(R^* \cap T) + 3 \gamma Q^*$.
\end{lemma}
\begin{proof}
By Eq.(\ref{eq04}), we reject all the jobs with fractional solutions and thus $P(A_t)$ will not exceed the optimal objective value of LP(\ref{eq03}).
Recall that setting $x_j = 1$ only for each $J_j \in R^* \cap T$ is a feasible solution for LP(\ref{eq03})
with objective value $P(A^* \cap T)$.
Therefore, we have $P(A_t) \le P(A^* \cap T)$.
The inequality $p_{\max}(A_t) \le p_{\max}(A^*)$ is straightforward by the first constraint of LP(\ref{eq03}).

From the third constraint of LP(\ref{eq03}) and Definition~\ref{def05} that $E(T_1) \le (t^*+1) \cdot 2\gamma Q^* \le E(R^* \cap T) + 2\gamma Q^*$.
Since each job $J_j \in T$ is $\gamma$-tiny, by Definition~\ref{def02}, $e_j \le \gamma Q^*$.
By Lemma~\ref{lemma06}, we conclude that 
\[
E(R_t) = E(T_1)+E(T_2) \le E(R^* \cap T) + 2\gamma Q^* + \gamma Q^* = E(R^* \cap T) + 3\gamma Q^*.
\]
The lemma is proved.
\end{proof}

\subsection{The complete $(2+\epsilon)$-approximation algorithm}

Using Operations~\ref{op01} and~\ref{op02}, we have determined the rejection decisions for all jobs.
That is, the jobs in $A_c \cup A_t$ are accepted while the jobs in $R_c \cup R_t$ are rejected.
By Lemmas~\ref{lemma05} and~\ref{lemma07}, we conclude that $P(A_c \cup A_t) \le P(A^*)$ and $p_{\max}(A_c \cup A_t) \le p_{\max}(A^*)$.
Therefore, by Lemma~\ref{lemma02}, there exists a schedule for $A_c \cup A_t$ such that the energy consumption constraint is satisfied and its makespan is at most $2C_{\max}(\pi^*)$.
Then, we can apply the PTAS $\Phi$ in Definition~\ref{def01} to the job set $A_c \cup A_t$ using $\frac \epsilon4$ as input and output the schedule $\Phi(A_c \cup A_t, \frac \epsilon4)$.
The complete algorithm is summarized in Algorithm~\ref{Approx1}.

\begin{algorithm}
\caption{Algorithm {\sc Approx1} for $P \mid rej, U, r_i  \mid C_{\max}(A) + \sum_{J_j \in R} e_j$}
\label{Approx1}
\begin{algorithmic}[1]
\State Input: A set $J$ of $n$ jobs and $m$ machines;

\State Guess the values $p_{\max}(A^*)$, $e_{\max}(R^*)$ and define the quantity $Q^*$ in Eq.(\ref{eq01}).

\State Set up a constant $\gamma = \frac \epsilon6$ and partition the jobs into $\gamma$-critical and $\gamma$-tiny as in Definition~\ref{def02}.

\State Use Operations~\ref{op01} and~\ref{op02} to determine the rejection decisions for the $\gamma$-critical and $\gamma$-tiny jobs, respectively.

\State Reject the jobs in $R_c \cup R_t$ and accept the jobs in $A_c \cup A_t$.
Output $\Phi(A_c \cup A_t, \frac \epsilon 4)$ as a schedule for $A_c \cup A_t$.
\end{algorithmic}
\end{algorithm}

\begin{theorem}
\label{thm01}
Algorithm~\ref{Approx1} is a $(2+\epsilon)$-approximation algorithm for $P \mid rej, U, r_i  \mid C_{\max}(A) + \sum_{J_j \in R} e_j$.
\end{theorem}
\begin{proof}
The dominant step in Algorithm~\ref{Approx1} is to guess the values of $c^*_i$ in Operation~\ref{op01}.
One can verify that the running time of Algorithm~\ref{Approx1} is polynomial in $n$ for any fixed $\epsilon$.

Next, we prove that Algorithm~\ref{Approx1} is a $(2+\epsilon)$-approximation.
By Eq.(\ref{eq01}) and Lemmas~\ref{lemma05}, \ref{lemma07}, we have 
\[
E(R_c \cup R_t) \le (1+\frac \epsilon 4) E(R^* \cap C) + E(R^* \cap T) + 3 \gamma Q^* 
\le (1+\frac \epsilon 4) E(R^*) + 3 \gamma Z^*.
\]
Using Lemma~\ref{lemma02}, $\Phi(A_c \cup A_t, \frac \epsilon4)$ satisfies the energy consumption constraint
and 
the makespan is at most $2(1+\frac \epsilon 4) C_{\max}(\pi^*)$.
Recall that $Z^* = E(R^*) + C_{\max}(\pi^*)$.
Therefore, by $\epsilon < 1$ and Eq.(\ref{eq01}), the total cost is at most 
\[
2(1+\frac \epsilon 4) C_{\max}(\pi^*) + (1+\frac \epsilon 4) E(R^*) + 3 \gamma Z^* \le (2+\frac \epsilon2) Z^* + \frac \epsilon 2 Z^* = (2+\epsilon)Z^*.
\]
The theorem is proved.
\end{proof}

\section{A QPTAS for $P \mid rej, U, r_i  \mid C_{\max}(A) + \sum_{J_j \in R} e_j$}

In this section, we present a QPTAS for $P \mid rej, U, r_i  \mid C_{\max}(A) + \sum_{J_j \in R} e_j$.
Let $\gamma = \frac {3\epsilon}{4n}$.
Our QPTAS first reject all $\gamma$-tiny jobs and then use Operation~\ref{op01} to determine the rejection decisions for the $\gamma$-critical jobs.
Therefore, the set of rejected jobs is $R_c \cup T$ while the jobs in $A_c$ are accepted.

\begin{lemma}
\label{lemma08}
$E(R_c \cup T) \le (1+\frac \epsilon4) E(R^*) + \frac {3\epsilon Z^*}4$.
There exists a schedule for $A_c$ such that the energy consumption constraint is satisfied and the makespan is at most $C_{\max}(\pi^*)$.
\end{lemma}
\begin{proof}
Note that $\gamma = \frac {3\epsilon}{4n}$ and each job in $T$ has a processing time at most $\gamma Q^*$.
So, $E(T) \le n \cdot \gamma Q^* = \frac {3\epsilon Q^*}4$.
By $R^* \cap C \subseteq R^*$, Lemma~\ref{lemma05} and Eq.(\ref{eq01}), we have
\[
E(R_c \cup T) = E(R_c) + E(T) \le (1+\frac \epsilon4) E(R^* \cap C) + \frac {3\epsilon Q^*}4 \le (1+\frac \epsilon4) E(R^*) + \frac {3\epsilon Z^*}4.
\]
The first inequality holds.

By Lemma~\ref{lemma04}, $A^* \cap C$ covers $A_c$ and by Definition~\ref{def04}, there exists an injective function mapping each job $J_j \in A_c$ to a job $J_{f(j)} \in A^* \cap C$ such that $p_{f(j)} \ge p_j$.
So, we can construct a schedule $\pi$ for $A_c$ by scheduling $J_j \in A_c$ on the same machine where $J_{f(j)}$ is assigned in $\pi^*$.
Since $p_{f(j)} \ge p_j$ and $f$ is injective, the load of each machine $M_i$ in $\pi$ is at most that of $M_i$ in $\pi^*$.
Clearly, $\pi$ is the desired schedule for $A_c$ and we are done.
\end{proof}

By Lemma~\ref{lemma08}, we can apply the PTAS $\Phi$ in Definition~\ref{def01} to schedule the jobs in $A_c$.
The algorithm is summarized as follows.

\begin{algorithm}
\caption{Algorithm {\sc Approx2} for $P \mid rej, U, r_i  \mid C_{\max}(A) + \sum_{J_j \in R} e_j$}
\label{Approx2}
\begin{algorithmic}[1]
\State Input: A set $J$ of $n$ jobs and $m$ machines;

\State Guess the values $p_{\max}(A^*)$, $e_{\max}(R^*)$ and define the quantity $Q^*$ in Eq.(\ref{eq01}).

\State Set up a constant $\gamma = \frac {3\epsilon}{4n}$ and partition the jobs into $\gamma$-critical and $\gamma$-tiny as in Definition~\ref{def02}.

\State Use Operation~\ref{op01} to determine the rejection decisions for the $\gamma$-critical jobs.

\State Reject the jobs in $R_c \cup T$ and accept the jobs in $A_c$.
Output $\Phi(A_c, \frac \epsilon 4)$ as a schedule for $A_c$.
\end{algorithmic}
\end{algorithm}

\begin{theorem}
\label{thm02}
Algorithm~\ref{Approx2} is a QPTAS for $P \mid rej, U, r_i  \mid C_{\max}(A) + \sum_{J_j \in R} e_j$.
\end{theorem}
\begin{proof}
The dominant step in Algorithm~\ref{Approx2} is to guess the values of $c^*_i$ in Operation~\ref{op01},
which takes $(n+1)^{O(\frac {\log 1/\gamma}\epsilon)}$ time.
Since $\gamma = \frac {3\epsilon}{4n}$, the time satisfies the requirement of a QPTAS.

Next, we prove the approximation ratio of Algorithm~\ref{Approx2}.
By Lemma~\ref{lemma08}, $\Phi(A_c, \frac \epsilon4)$ satisfies the energy consumption constraint
and 
the makespan is at most $(1+\frac \epsilon 4) C_{\max}(\pi^*)$.
Therefore, by Eq.(\ref{eq01}) and Lemma~\ref{lemma08}, the total cost is at most 
$(1+\frac \epsilon4)(C_{\max}(\pi^*)+E(R^*)) + \frac {3\epsilon Z^*}4 = (1+\epsilon) Z^*$.
The theorem is proved.
\end{proof}

\section{A PTAS for $P \mid rej, U, r_i=r  \mid C_{\max}(A) + \sum_{J_j \in R} e_j$}

In this section, we assume the energy consumption rates are uniform, that is, $r_i = r$ for every $i$.
Therefore, the energy consumption constraint is indeed saying that the total processing time of the accepted jobs is at most $U/r$.
We set up two constants $\beta, \gamma$ as follows.
\begin{equation}
\label{eq05}
\beta = \frac \epsilon6 \mbox{ and } \gamma=\beta^3 = \frac {\epsilon^3}{216}.
\end{equation}
Then, Operation~\ref{op01} can determine the rejection decisions for the $\gamma$-critical jobs in $(n+1)^{O(1/\epsilon^4)}$ time.
Then, we partition the $\gamma$-tiny jobs into {\em long} and {\em short} based on $\beta$.

\begin{definition}
\label{def06}
A $\gamma$-tiny job $J_j$ is {\em long} if $p_j > \beta Q^*$; otherwise, $J_j$ is {\em short}.
Let $L$ and $S$ be the set of long and short jobs, respectively.
Therefore, $T = L \cup S$.
\end{definition}

We let $L_0 = \{ J_j \in L: p_j > Q^* \}$.
Note that for a job $J_j \in L \setminus L_0$, there exists an integer $\frac 1\beta \le i \le \frac 1{\beta^2}-1$ 
such that $i \beta^2 Q^* < p_j \le (i+1) \beta^2 Q^*$.
So, the long jobs in $L \setminus L_0$ can be partitioned into the following subgroups.
\begin{equation}
\label{eq06}
L_i = \left \{ J_j \in L \setminus L_0: i \cdot \beta^2 Q^* < p_j \le (i+1) \beta^2 Q^* \right \},
\mbox{ for each integer $\frac 1\beta \le i \le \frac 1{\beta^2}-1$.}
\end{equation}

\begin{lemma}
\label{lemma09}
The number of subgroups $L_i$ is at most $O(\frac 1{\epsilon^2})$.
Moreover, if $J_j, J_k \in L_i$ for some $\frac 1\beta \le i \le \frac 1{\beta^2}-1$, then $p_j \le (1+\beta)p_k$.
\end{lemma}
\begin{proof}
The total number of subgroups $L_i$ is at most $\frac 1{\beta^2}-\frac 1\beta+1 = O(\frac 1{\epsilon^2})$.
If $J_j, J_k \in L_i$, then by Eq.(\ref{eq06}), we have $p_j - p_k \le \beta^2 Q^*$.
Since $J_k$ is long, $p_k > \beta Q^*$ and thus, $p_j - p_k \le \beta^2 Q^* < \beta p_k$.
The proof is completed.
\end{proof}

The following definition is presented for describing the number of jobs in each $L_i$ accepted in the optimal solution.

\begin{definition}
\label{def07}
Let $\ell^*_i$ be the number of jobs in $L_i$ that are accepted in the optimal solution for every non-negative integer $i$.
Note that by Eq.(\ref{eq01}), $p_j > Q^* \ge p_{\max}(A^*)$ for each job $J_j \in L_0$ and thus, $\ell^*_0 = 0$.
We know $\ell^*_i \in \{ 0, \ldots, n \}$ for each positive integer $i$
and by Lemma~\ref{lemma09}, the values of $\ell^*_i$ can be guessed in $(n+1)^{O(1/\epsilon^2)}$, which is polynomial in $n$ for any fixed $\epsilon$.
\end{definition}
We continue to use the value $t^* = \lfloor \frac {E(R^* \cap T)}{2\gamma Q^*} \rfloor$ in Definition~\ref{def05},
which can be guessed in $O(\frac n{\epsilon^3})$ time.
Similarly to LP(\ref{eq03}), we present a linear program LP(\ref{eq03}) to make rejection decisions for all $\gamma$-tiny jobs.
For each $\gamma$-tiny job $J_j \in T$, we let $x_j \in \{ 0, 1 \}$ be a binary decision variable such that $x_j = 1$ if and only if $J_j$ is rejected in the optimal solution 
and $y_j = 1-x_j$.
We relax the variables $x_j, y_j$ to be continuous.
One sees that LP(\ref{eq07}) minimizes the total processing time of the accepted jobs in $T$
while the number of accepted jobs in $L_i$ is exactly $\ell^*_i$ and the total penalty of the rejected jobs in $T$ is bounded by  $(t^*+1) \cdot 2\gamma Q^*$.

\begin{equation}
\label{eq07}
\begin{aligned}
 \min \quad & \sum_{J_j \in T} y_j p_j   \\
& x_j + y_j = 1, \quad  \forall J_j  \in T \\
& \sum_{J_j \in L_i} y_j = \ell^*_i, \quad  \forall i = 0 \mbox{ or } \frac 1\beta \le i \le \frac 1{\beta^2}-1 \\
& \sum_{J_j \in T} x_j e_j  \le (t^*+1)\cdot 2\gamma Q^*.
\end{aligned}
\end{equation}
LP(\ref{eq07}) has a feasible solution if we set $x_j = 1$ if and only if $J_j \in R^* \cap T$.
So, we can compute a basic solution for LP(\ref{eq07}) and partition the $\gamma$-tiny jobs into three sets as in Eq.(\ref{eq04}).
Let $R_t = T_1 \cup T_2$ and a $\gamma$-tiny job $J_j$ is accepted if and only if $J_j \in A_t$.
The operation is summarized below.

\begin{operation}
\label{op03}
Given $\gamma, \beta$ in Eq.(\ref{eq05}), we first partition the $\gamma$-tiny jobs into $L_0$ and the subgroups $L_i$ as in Eq.(\ref{eq06}).
For each $L_i$, we guess the number $\ell^*_i$ as defined in Definition~\ref{def07} in $(n+1)^{O(1/\epsilon^2)}$ time.

Then, we guess the value $t^*$ as defined in Definition~\ref{def05} and establish the linear program LP(\ref{eq07}).
Subsequently, we compute a basic solution and partition the $\gamma$-tiny jobs into $A_t$ and $R_t = T_1 \cup T_2$ as defined in Eq.(\ref{eq04}).
The jobs in $A_t$ are accepted while the jobs in $R_t = T_1 \cup T_2$ are rejected.
\end{operation}

Recall that $\frac 1\beta$ is an integer.
We show similar results as in Lemmas~\ref{lemma06} and~\ref{lemma07}.

\begin{lemma}
\label{lemma10}
$|T_2| \le \frac 1{\beta^2}-2$.
\end{lemma}
\begin{proof}
By $\epsilon \le 1$, we have $\frac 1\beta = \frac 6\epsilon \ge 6$.
So, the number of constraints in LP(\ref{eq07}) is $|T| + \frac 1{\beta^2}-\frac 1\beta+2 < |T|+\frac 1{\beta^2}-2$.
The number of positive values in the basic solution is at most $|T|+\frac 1{\beta^2}-2$.
Clearly, the number of positive values is exactly $|A_t| + |T_1| + 2|T_2| = |T| + |T_2| \le |T| + \frac 1{\beta^2}-2$.
It follows that $|T_2| \le \frac 1{\beta^2}-2$ and the lemma is proved.
\end{proof}

\begin{lemma}
\label{lemma11}
$|A_t \cap L_i| \le |A^* \cap L_i|$ for every $i = 0$ or $\frac 1\beta \le i \le \frac 1{\beta^2}-1$
and $P(A_t) \le P(A^* \cap T)$.
Moreover, $E(R_t) \le E(R^* \cap T) + \beta Q^*$.
\end{lemma}
\begin{proof}
Recall that $|A^* \cap L_i| = \ell^*_i$.
So, the inequality $|A_t \cap L_i| \le |A^* \cap L_i|$ comes from the second constraint of LP(\ref{eq07}).
Similarly to Lemma~\ref{lemma07}, the second inequality $P(A_t) \le P(A^* \cap T)$ can be proved.
Moreover, $E(T_1) \le E(R^* \cap T) + 2\gamma Q^*$ and $e_j \le \gamma Q^*$ for each $J_j \in T_2$.
By Lemma~\ref{lemma10} and Eq.(\ref{eq05}), we conclude that 
\[
E(R_t) = E(T_1)+E(T_2) \le E(R^* \cap T) + 2\gamma Q^* + (\frac 1{\beta^2} - 2) \gamma Q^* = E(R^* \cap T) + \beta Q^*
\]
The lemma is proved.
\end{proof}

Recall that the jobs in $A_c \cup A_t$ are accepted and we will show that there exists a schedule for $A_c \cup A_t$
such that the energy consumption constraint is satisfied and the makespan is at most $(1+\beta)C_{\max}(\pi^*)+\beta Q^*$.

\begin{lemma}
\label{lemma12}
There exists a schedule for $A_c \cup A_t$
such that the energy consumption constraint is satisfied and the makespan is at most $(1+\beta)C_{\max}(\pi^*)+\beta Q^*$.
\end{lemma}
\begin{proof}
By Lemmas~\ref{lemma05} and~\ref{lemma11}, we have $P(A_c \cup A_t) \le P(A^* \cap C) + P(A^* \cap T) = P(A^*) \le U/r$.
The energy consumption constraint is always satisfied.
We construct a schedule $\pi$ for $A_c \cup A_t$ as follows:
\begin{itemize}
\item[1.] By Lemma~\ref{lemma04} and Definition~\ref{def04}, there is an injective function $f$ from $A_c$ to $A^* \cap C$ such that $p_{f(j)} \ge p_j$.
We schedule a job $J_j \in A_c$ on the same machine where $J_{f(j)}$ is assigned in the schedule $\pi^*$.

\item[2.] Given a machine, we greedily schedule the jobs in $A_t \cap L_i$ on the machine until the number of assigned jobs is exactly the same as that of $\pi^*$
or we run out of jobs.
By Lemma~\ref{lemma11}, the jobs in $A_t \cap L_i$ have finished their assignments.

\item[3.] Finally, we greedily schedule the jobs in $A_t \cap S$ on the least load machine.
\end{itemize}

It suffices to show that the makespan of $\pi$ is at most $(1+\beta)C_{\max}(\pi^*)+\beta Q^*$.
If the makespan of $\pi$ is determined by a job in $A_t \cap S$, then 
since the processing time of a short job is at most $\beta Q^*$ and the job is scheduled on a least load machine, 
the makespan is at most $\frac 1m P(A_c \cup A_t) + \beta Q^*$.
Recall that $P(A_c \cup A_t) \le P(A^*)$ and $C_{\max}(\pi^*) \ge \frac 1m P(A^*)$, the makespan of $\pi$ is at most $C_{\max}(\pi^*)+\beta Q^*$.

Otherwise, the makespan of $\pi$ is determined by a job in $A_c$ or some $L_i$.
Let $M_s$ be a machine that achieves the makespan.
Since the jobs in $A_t \cap S$ are scheduled last, every job scheduled on $M_s$ belongs to $A_c$ or some $L_i$.
By Lemma~\ref{lemma11} and Definition~\ref{def07}, $|A_t \cap L_0| = 0$. 
By the first item in the construction of $\pi$, it suffices to consider the jobs in $\cup_{i=1/\beta}^{1/\beta^2-1} L_i$.
For each $i$, the number of jobs in $L_i$ scheduled on the machine $M_s$ in $\pi$ is no more than that of $M_s$ in $\pi^*$.
By Lemma~\ref{lemma09}, the makespan of $\pi$ is at most $(1+\beta)C_{\max}(\pi^*)$.
The proof is completed.
\end{proof}

By Lemma~\ref{lemma12}, we can apply the PTAS $\Phi$ to schedule the jobs in $A_c \cup A_t$ 
and output the schedule $\Phi(A_c \cap A_t, \frac \epsilon6)$.
The algorithm is presented in Algorithm~\ref{Approx3}.

\begin{algorithm}
\caption{Algorithm {\sc Approx3} for $P \mid rej, U, r_i=r  \mid C_{\max}(A) + \sum_{J_j \in R} e_j$}
\label{Approx3}
\begin{algorithmic}[1]
\State Input: A set $J$ of $n$ jobs and $m$ machines;

\State Guess the values $p_{\max}(A^*)$, $e_{\max}(R^*)$ and define the quantity $Q^*$ in Eq.(\ref{eq01}).

\State Set up two constants $\gamma, \beta$ as in Eq.(\ref{eq05}) and partition the jobs into $\gamma$-critical and $\gamma$-tiny as in Definition~\ref{def02}.

\State Use Operations~\ref{op01} and~\ref{op03} to determine the rejection decisions for the $\gamma$-critical and $\gamma$-tiny jobs, respectively.

\State Reject the jobs in $R_c \cup R_t$ and accept the jobs in $A_c \cup A_t$.
Output $\Phi(A_c \cup A_t, \frac \epsilon 6)$ as a schedule for $A_c \cup A_t$.
\end{algorithmic}
\end{algorithm}

\begin{theorem}
\label{thm03}
Algorithm~\ref{Approx3} is a PTAS for $P \mid rej, U, r_i = r \mid C_{\max}(A) + \sum_{J_j \in R} e_j$.
\end{theorem}
\begin{proof}
The dominant step in Algorithm~\ref{Approx3} is to guess the values of $c^*_i, \ell^*_i$ in Operations~\ref{op01} and~\ref{op03}.
One can verify the total running time is polynomial in $n$ for a fixed $\epsilon$.

Next, we prove that Algorithm~\ref{Approx3} is a $(1+\epsilon)$-approximation.
By Lemma~\ref{lemma12}, the schedule $\Phi(A_c \cup A_t, \frac \epsilon6)$ satisfies the energy consumption constraint
and 
by $\beta = \frac \epsilon6$, $\epsilon < 1$ and Eq.(\ref{eq01}),
the makespan is at most 
\[
(1+\frac \epsilon 6)^2 C_{\max}(\pi^*) + (1+\frac \epsilon6) \beta Q^* 
\le (1+\frac \epsilon 6)^2 C_{\max}(\pi^*) + 2\beta Z^* \le (1+\frac \epsilon2)C_{\max}(\pi^*)+2\beta Z^*.
\]
By Eq.(\ref{eq01}) and Lemmas~\ref{lemma05}, \ref{lemma11}, we have 
\[
E(R_c \cup R_t) \le (1+\frac \epsilon 4) E(R^* \cap C) + E(R^* \cap T) + \beta Q^* 
\le (1+\frac \epsilon 4) E(R^*) + \beta Z^*.
\]
Therefore, by $\beta = \frac \epsilon6$ again, the total cost is at most 
$(1+\frac \epsilon2)(C_{\max}(\pi^*)+E(R^*)) + 3\beta Z^* = (1+\epsilon) Z^*$.
The theorem is proved.
\end{proof}

\section{An FPTAS for $Pm \mid rej, U, r_i  \mid C_{\max}(A) + \sum_{J_j \in R} e_j$}

In this section, we present an FPTAS for the problem $Pm \mid rej, U, r_i  \mid C_{\max}(A) + \sum_{J_j \in R} e_j$ in which the number $m$ of machines is a fixed constant.
Without loss of generality, we assume all processing times $p_j$ and rejection penalties $e_j$ are positive integers.
By Eq.(\ref{eq01}), $E(R^*) \le n \cdot Q^*$ and the load of each machine in a solution is at most $n \cdot Q^*$.
Given a solution for the partial job set $\{ J_1, \ldots, J_j \}$, we let $\vec{v} = (v_1, \ldots, v_{m+1})$ be an $(m+1)$-dimensional vector 
where $v_i$ stores the load of $M_i$, $i = 1, \ldots, m$ and $v_{m+1}$ stores the total penalty of the rejected jobs.
Note that $v_i \in \{ 0, \ldots, n Q^* \}$ for each $i \le m+1$.
So, there are at most $O((nQ^*)^{m+1})$ distinct vectors for $\{ J_1, \ldots, J_j \}$.

Let $V_0 = \{ \vec{0} \}$ and $V_j$ be the vector space containing all vectors for $\{ J_1, \ldots, J_j \}$.
We present a dynamic programming (DP1 for short) to compute the space $V_j$ from $V_{j-1}$ below.

\begin{algorithm}
\caption{DP1 for computing the vector space $V_j$}
\label{DP1}
\begin{algorithmic}[1]
\State Input: the vector space $V_{j-1}$; 

\State Initially $V_j = \{ \vec{0} \}$.

\State If $p_j > Q^*$, then $J_j$ must be rejected. 
We add the vector $(v_1, \ldots, v_m, v_{m+1}+e_j)$ to $V_j$.

\State If $e_j > Q^*$, then $J_j$ must be accepted and scheduled on some machine $M_i$.
For every $i=1, \ldots, m$, we add the vector $(v_1, \ldots, v_i+p_j, \ldots, v_m, v_{m+1})$ to $V_j$.

\State If $p_j \le Q^*$ and $e_j \le Q^*$, then we add the vector $(v_1, \ldots, v_m, v_{m+1}+e_j)$ and for every $i=1, \ldots, m$, we add the vector $(v_1, \ldots, v_i+p_j, \ldots, v_m, v_{m+1})$ to $V_j$.

\State If there exist multiple identical vectors in $V_j$, then we keep only one vector.
\end{algorithmic}
\end{algorithm}

Recall that $|V_{j-1}| \le O((nQ^*)^{m+1})$.
By the above DP1, we can construct the vector space $V_j$ in $O(m(nQ^*)^{m+1})$ time but $|V_j| \le O((nQ^*)^{m+1})$ still holds.
Therefore, the final vector space $V_n$ can be constructed in $O(mn(nQ^*)^{m+1})$ time.
For each $\vec{v} \in V_n$, the energy consumption is $\sum_{i=1}^m r_i v_i$.
So, we can check every vector in $V_n$ and output a vector $\vec{v}$ with the minimum $\max_{i=1}^m v_i + v_{m+1}$
subject to the energy consumption constraint $\sum_{i=1}^m r_i v_i \le U$.

Note that the running time of DP1 is $O(mn(nQ^*)^{m+1})$, which is pseudo-polynomial when $m$ is a fixed constant.
Next, we apply the ``vector trimming'' technique in~\cite{W00} to design an FPTAS.
Given a fixed constant $\epsilon \le 1$, let $\mu = \frac {\epsilon Q^*}{2n}$.
Since each vector in each $V_j$ must fall in the $(m+1)$-cube $[0, nQ^*]^{m+1}$, we divide the $(m+1)$-cube into {\em boxes} of side length $\mu$.

\begin{definition}
\label{def08}
A {\em box} is an $(m+1)$-tuple $(b_1, \ldots, b_{m+1})$ where each $b_i \in \{ 1, \ldots, \frac {2n^2}\epsilon, \frac {2n^2}\epsilon+1 \}$.
A vector $\vec{v}=(v_1, \ldots, v_{m+1})$ falls in the box $(b_1, \ldots, b_{m+1})$ 
if $(b_i-1) \mu \le v_i < b_i \cdot \mu$ for every $i = 1, \ldots, m+1$.
So, the $(m+1)$-cube $[0, nQ^*]^{m+1}$ are divided into $O(\frac {2^m n^{2m}}{\epsilon^m})$ distinct boxes.
\end{definition}
We let $V^\#_j$ be the trimmed space of $V_j$ such that at most one vector in $V^\#_j$ falls in each box.
Initially, we set $V^\#_0 = V_0 = \{ \vec{0} \}$.
Similarly to DP1, we present a dynamic programming DP2 to construct $V^\#_j$ from $V^\#_{j-1}$ as follows:
We first call DP1 in Algorithm~\ref{DP1} by using $V^\#_{j-1}$ as input and the vector space outputted is denoted by $V'_j$.
If multiple vectors in $V'_j$ fall into a same box, then we choose only one vector with the minimum energy consumption.
The remaining vectors of $V'_j$ form the trimmed space $V^\#_j$.
The DP2 outputs a vector $\vec{v}$ in $V^\#_n$ with the minimum $\max_{i=1}^m v_i + v_{m+1}$
subject to the energy consumption constraint $\sum_{i=1}^m r_i v_i \le U$.

Note that for every $j = 1, \ldots, n$, the number of vectors in the trimmed space $V^\#_j$ is bounded by the number of boxes, that is, $O(\frac {2^m n^{2m}}{\epsilon^m})$.
Therefore, it takes at most $O(\frac {m 2^m n^{2m}}{\epsilon^m})$ time to construct $V'_j$ and $V^\#_j$.
So, the total running time of DP2 is at most $O(\frac {m 2^m n^{2m+1}}{\epsilon^m})$, which is polynomial in $n$ and $\frac 1\epsilon$ when $m$ is a fixed constant.
The following lemma is vital to show the approximation ratio.

\begin{lemma}
\label{lemma13}
For each vector $\vec{v} = (v_1, \ldots, v_{m+1})$ in some $V_j$, there exists a vector $\vec{v}^\# = (v^\#_1, \ldots, v^\#_{m+1})$ in $V^\#_j$ such that 
(1) the total energy consumption of $\vec{v}^\#$ is no more than that of $\vec{v}$ and;
(2) for every $i = 1, \ldots, m+1$, we have $v^\#_i \le v_i + j \mu$.
\end{lemma}
\begin{proof}
We prove the lemma by induction.
Clearly, when $j = 0$, $V_0 = V^\#_0 = \{ \vec{0} \}$ and the lemma holds.
Then we assume the lemma holds for $j-1$.
For ease of presentation, given a vector $\vec{v}$, we let $EC(\vec{v})$ be the total energy consumption of $\vec{v}$.

Consider a vector $\vec{v} = (v_1, \ldots, v_{m+1}) \in V_j$, which is generated by the vector $\vec{u} = (u_1, \ldots, u_{m+1}) \in V_{j-1}$.
By the inductive hypothesis, there exists a vector $\vec{u}^\# = (u^\#_1, \ldots, u^\#_{m+1}) \in V^\#_{j-1}$
such that $EC(\vec{u}^\#) \le EC(\vec{u})$ and;
for every $i = 1, \ldots, m+1$, we have $u^\#_i \le u_i + (j-1) \mu$.
Note that when constructing $\vec{v}$, $J_j$ is rejected or (accepted and) scheduled on some machine $M_i$.

{\em Case 1.} $J_j$ is rejected. 
Then $v_s = u_s$ for each $s \le m$ and $v_{m+1}=u_{m+1}+e_j$.
Therefore, $EC(\vec{v}) = EC(\vec{u})$.
In this case, $\vec{u}' = (u^\#_1, \ldots, u^\#_{m+1}+e_j)$ must appear in $V'_j$ when DP2 considers $\vec{u}^\#$.
By the trimming process, there exists a vector $\vec{v}^\# = (v^\#_1, \ldots, v^\#_{m+1})$ in $V^\#_j$ 
such that $\vec{v}^\#$ and $\vec{u}'$ fall in the same box and $EC(\vec{v}^\#) \le E(\vec{u}')$.
Note that $EC(\vec{u}') = EC(\vec{u}^\#) \le EC(\vec{u}) = EC(\vec{v})$.
We conclude that $EC(\vec{v}^\#) \le EC(\vec{v})$.

Note that if two vectors falls in the same box, then the diffence in each coordinate is at most $\mu$.
Moreover, $u^\#_s \le u_s + (j-1) \mu$ for every $s \le m+1$ (see the second paragraph in the proof).
Consequently, for each $s \le m$, we have $v^\#_s \le u^\#_s + \mu \le u_s + j\mu$ and $v^\#_{m+1} \le u^\#_{m+1}+e_j+\mu \le u_{m+1}+e_j + j\mu$.
It follows from $v_s = u_s$ for each $s \le m$ and $v_{m+1}=u_{m+1}+e_j$ that $v^\#_s \le v_s + j\mu$ for every $s \le m+1$.
So, the vector $\vec{v}^\#$ is as desired.

{\em Case 2.}  $J_j$ is accepted and scheduled on some machine $M_i$.
In this case, $v_s = u_s$ for each $s \ne i$ and $v_i = u_i + p_j$.
Clearly, $EC(\vec{v}) = EC(\vec{u})+r_i p_j$.
When DP2 considers $\vec{u}^\#$, the vector $\vec{u}' = (u^\#_1, \ldots, u^\#_i+p_j, \ldots, u^\#_{m+1})$ will be added to $V'_j$.
Similarly to Case 1, there exists a vector $\vec{v}^\# = (v^\#_1, \ldots, v^\#_{m+1})$ in $V^\#_j$ 
such that $\vec{v}^\#$ and $\vec{u}'$ fall in the same box and $EC(\vec{v}^\#) \le EC(\vec{u}')$.
By $EC(\vec{u}^\#) \le EC(\vec{u})$ (see the second paragraph of the proof), we have $EC(\vec{u}') = EC(\vec{u}^\#)+r_i p_j \le EC(\vec{u})+r_i p_j$.
It follows from $EC(\vec{v}) = EC(\vec{u})+r_i p_j$ that $EC(\vec{v}^\#) \le EC(\vec{u})+r_i p_j = EC(\vec{v})$.

Since $\vec{v}^\#$ and $\vec{u}'$ are in the same box and $u^\#_s \le u_s + (j-1) \mu$ for every $s \le m+1$ (see the second paragraph of the proof), 
we know $v^\#_i \le u^\#_i+p_j+\mu \le u_i+p_j+j\mu$
and 
for each $s \ne i$, we have $v^\#_s \le u^\#_s+\mu \le u_s+j\mu$.
Recall that $v_s = u_s$ for each $s \ne i$ and $v_i = u_i + p_j$.
We can conclude that $v^\#_s \le v_s+j\mu$ for every $s \le m+1$ and the lemma is proved.
\end{proof}

\begin{theorem}
\label{thm04}
DP2 is an FPTAS for $Pm \mid rej, U, r_i  \mid C_{\max}(A) + \sum_{J_j \in R} e_j$.
\end{theorem}
\begin{proof}
Recall that the running time of DP2 is $O(\frac {m 2^m n^{2m+1}}{\epsilon^m})$ and it suffices to show the approximation ratio of DP2.
Note that there exists a vector $\vec{v} \in V_n$ such that its energy consumption is at most $U$ and the cost is $Z^* = \max_{i=1}^m v_i + v_{m+1}$.
By Lemma~\ref{lemma13}, there is vector $\vec{v}^\# \in V^\#_n$ that the energy consumption is at most that of $\vec{v}$
and by Eq.(\ref{eq01}), its cost is bounded by 
\[
\max_{i=1}^m v^\#_i + v^\#_{m+1} \le \max_{i=1}^m v_i + v_{m+1} + 2n\mu = Z^*+ \epsilon Q^* \le (1+\epsilon)Z^*.
\]
The proof is finished.
\end{proof}

\section{Conclusions}

We study the scheduling with rejection problem $P \mid rej, U, r_i \mid C_{\max}(A)+\sum_{J_j \in R} e_j$ in this paper,
where the energy consumption rates may be non-uniform.
We present the first $(2+\epsilon)$-approximation algorithm and a QPTAS for the problem as well as a PTAS when the energy consumption rates are uniform.
When the number $m$ of machines is a fixed constant, we provide an FPTAS.

There are several possible future research directions.
First, it would be challenging and interesting to design a PTAS for $P \mid rej, U, r_i \mid C_{\max}(A)+\sum_{J_j \in R} e_j$.
Second, it would be possible to extend our model to more general settings.
For example, we can consider the related machines or assume that jobs have different release times.
Finally, it would be interesting to consider different objective values such as the total completion time of the accepted jobs plus the total rejection penalty of the rejected jobs.




\bibliography{main.bib}

\end{document}